\documentclass[12pt]{article}

\usepackage{times}
\usepackage{graphicx}
\usepackage{amsmath}
\usepackage{amssymb}
\usepackage{booktabs}
\usepackage{url}
\usepackage{hyperref}

\title{Thermal Noise as a Window into the Quantum Vacuum: Spatial Patterns Revealed by Simple Experiments}

\author{
   Sun-Hyun Youn$^{1}$  \\
    $^{1}$Department of Physics, Chonnan National University, Gwangju, Korea  \\
    {\small Email:sunyoun@jnu.ac.kr}
}

\date{\today} 

\begin{document}

\maketitle

\begin{abstract}
We show that the spatial structure of electromagnetic vacuum fluctuations, predicted by quantum electrodynamics, can be indirectly observed using thermal noise at radio frequencies. Using simple lab equipment like coaxial cables and RF splitters, we detect a clear suppression of thermal noise near conducting boundaries, mirroring the expected modulation of vacuum modes. This provides accessible, experimental evidence for quantum vacuum behavior without requiring advanced optics or cryogenics.

\end{abstract}

\section{Introduction}

The behavior of electromagnetic fields near a mirror surface is fundamentally altered by boundary conditions, leading to the formation of standing waves. This, in turn, modifies the vacuum fluctuations, impacting phenomena like spontaneous emission. While theoretical predictions suggest a reduction in vacuum fluctuations near a mirror, direct experimental verification in the optical regime has been challenging due to limitations in spatial resolution. Here, we propose and demonstrate an approach using radio-frequency (RF) measurements of thermal noise to indirectly probe the spatial distribution of vacuum fluctuations near a mirror. We leverage the fact that thermal noise, dominant in the RF domain, populates the same electromagnetic modes dictated by the boundary conditions as vacuum fluctuations. By characterizing the spatial distribution of thermal noise, we infer the corresponding modification of vacuum modes and, consequently, the vacuum fluctuations.

When light, an electromagnetic wave, is incident upon a mirror, the electric field must vanish at the mirror surface (assuming a perfect conductor). For a plane wave propagating in the $z$-direction and a mirror in the $xy$-plane, this boundary condition results in a standing wave pattern. The electric field intensity is zero at integer multiples of half the wavelength from the mirror and reaches a maximum at odd integer multiples of a quarter wavelength. This principle forms the basis of optical resonators (cavities), where only specific wavelengths (resonant modes) satisfying the boundary conditions imposed by two mirrors can exist.

It is well-established that this modification of the electromagnetic field also affects the vacuum fluctuations, the zero-point energy of the electromagnetic field \cite{casimir1948}. These vacuum fluctuations are linked to shot noise in optical measurements, and it has been demonstrated that they can be reduced below the standard quantum limit through nonlinear optical interactions \cite{jae1980}. In 1993, one of us theoretically predicted that the magnitude of vacuum fluctuations could be reduced near a mirror surface, simply due to the boundary conditions, without requiring nonlinear media \cite{youn1995}. While indirect evidence supporting this prediction was obtained \cite{wadood2019}, direct experimental verification in the optical domain has remained elusive due to the difficulty of achieving sub-wavelength spatial resolution in the direction of light propagation.

The fundamental properties of electromagnetic waves, including the modification of modes near a boundary, are not limited to the optical regime. We therefore propose an experimental approach in a different frequency domain, specifically the radio-frequency (RF) range, to overcome the limitations of optical measurements.

The vacuum fluctuations of a specific electromagnetic mode correspond to the lowest possible energy state of that mode, as defined by its frequency. To isolate and measure these vacuum fluctuations, one must eliminate all contributions from thermal noise and other external signals at the corresponding frequency. This condition can only be approached by cooling the system to absolute zero (0 K) and shielding it from any extrinsic electromagnetic interference. 

A compelling question then arises: once a state composed solely of vacuum fluctuations has been established, how does the subsequent introduction of thermal noise manifest spatially in terms of energy distribution? At a given frequency, the thermal energy added to the system acts to further excite the electromagnetic mode that initially harbored only the vacuum fluctuation energy. This added excitation effectively modifies the population of the mode, leading to a new energy distribution shaped by both vacuum and thermal contributions.

Now, consider a resonator formed by two mirrors, in thermal equilibrium with a reservoir.  The electromagnetic field inside the resonator will also reach thermal equilibrium, with energy distributed according to Planck's law.  However, crucially, the frequencies are no longer continuous but are discrete, corresponding to the resonant modes of the cavity.  The energy introduced by thermal noise will populate these discrete modes, adjusting the photon number in each mode to reach thermal equilibrium.

This key insight reveals that by measuring the spatial distribution of thermal noise at a specific frequency $f$, we can infer the spatial distribution of the electromagnetic mode at that frequency.  This, in turn, allows us to predict the spatial characteristics of the vacuum fluctuations at frequency $f$. In essence, the thermally added photons do not alter the spatial mode structure (which is determined by the boundary conditions), but only affect the excitation level (photon number) within those modes.

Therefore, by measuring the spatial distribution of RF thermal noise near a mirror, we can infer the spatial distribution of the vacuum modes near the mirror and, consequently, how the vacuum fluctuations are modified.

While direct measurement of vacuum fluctuations in the RF domain is challenging due to the overwhelming thermal noise, measuring the spatial variation of the thermal noise allows us to deduce how the vacuum modes are shaped near the mirror, providing indirect access to the behavior of vacuum fluctuations.  This approach opens a new avenue for experimentally investigating quantum electrodynamic effects near boundaries, circumventing the limitations of optical measurements.

Our experimental configuration utilizes BNC cables, with one end modified to achieve a zero impedance, effectively creating a mirror-like boundary condition. In a simplified scenario, an amplifier with a 50 $\Omega$ input impedance connected to a 50 $\Omega$ BNC cable of length $L$, terminated with a load impedance $Z_L$, would primarily detect thermal noise originating from the amplifier's input stage. The input impedance in this setup is well-described by transmission line theory, being a function of cable length and load impedance.

However, we interpret the observed thermal noise signal not merely as a passive measurement, but as a consequence of energy from the amplifier's input stage thermal noise exciting vacuum modes within the cable, leading to the formation of electromagnetic waves. Therefore, measuring the spatial characteristics of this thermally-induced noise within the cable is equivalent to probing the spatial characteristics of the vacuum modes themselves, effectively providing a measurement of the vacuum noise's spatial properties within the cable environment.

In this work, we first characterized the frequency-dependent thermal noise by varying both the length of the BNC cable connected to the amplifier input and the termination impedance ( 0 and infinity). While varying cable length for a fixed frequency is possible, we chose to analyze thermal noise across a range of frequencies for a fixed cable length to determine the spatial distribution of electromagnetic waves within the cable. This provided a comprehensive characterization of the electromagnetic mode structure.

Furthermore, analogous to how a beam splitter combines and separates optical beams, radio-frequency (RF) splitters can also manipulate electromagnetic waves through reflection and transmission. We employed a 4x4 RF splitter, connecting BNC cables to two input ports to define specific electromagnetic modes. By characterizing the propagation of these modes through the splitter, we investigated the behavior of vacuum fluctuations in the presence of a mirror-like boundary, analogous to studying vacuum fluctuations near a reflecting surface. This configuration allows us to probe how zero-point energy manifests in such a constrained RF environment.

\section{Spatial Mapping of Thermal Noise in Coaxially Constrained Geometries: Probing Vacuum Fluctuation Modes} \label{Methods1}

We consider a system comprising a voltage source ($v_b$) in series with an impedance ($Z_b$), connected to a transmission line of length $L_A$ and characteristic impedance $Z_0$. The transmission line is terminated with a load impedance $Z_l$. While the voltage across the device can be determined using equivalent impedance methods, we present here an analysis based on the theory of propagating electromagnetic waves\cite{pozar2011,shahmoon2017}.

The voltage originating from $v_b$ is initially divided at the input of the transmission line. Due to the series connection of $Z_b$  and $Z_0$, the voltage appearing at the input of the transmission line  ($v_0$) is given by:

\begin{eqnarray}
v_{0} =  v_b  \frac{Z_0 }{Z_b + Z_0}. 
\label{voltage0}
\end{eqnarray}

This voltage $v_0$ propagates along the transmission line of length $L_A$ to the load impedance $Z_l$.
At the load, a portion of the wave is reflected. The reflected wave travels back along the transmission line (length $L_A$) to the input of the device. The voltage of this first reflected wave arriving back at the input ($v_{1m}$) is:

\begin{figure}[h!]
    \centering
    \includegraphics[width=0.8\textwidth]{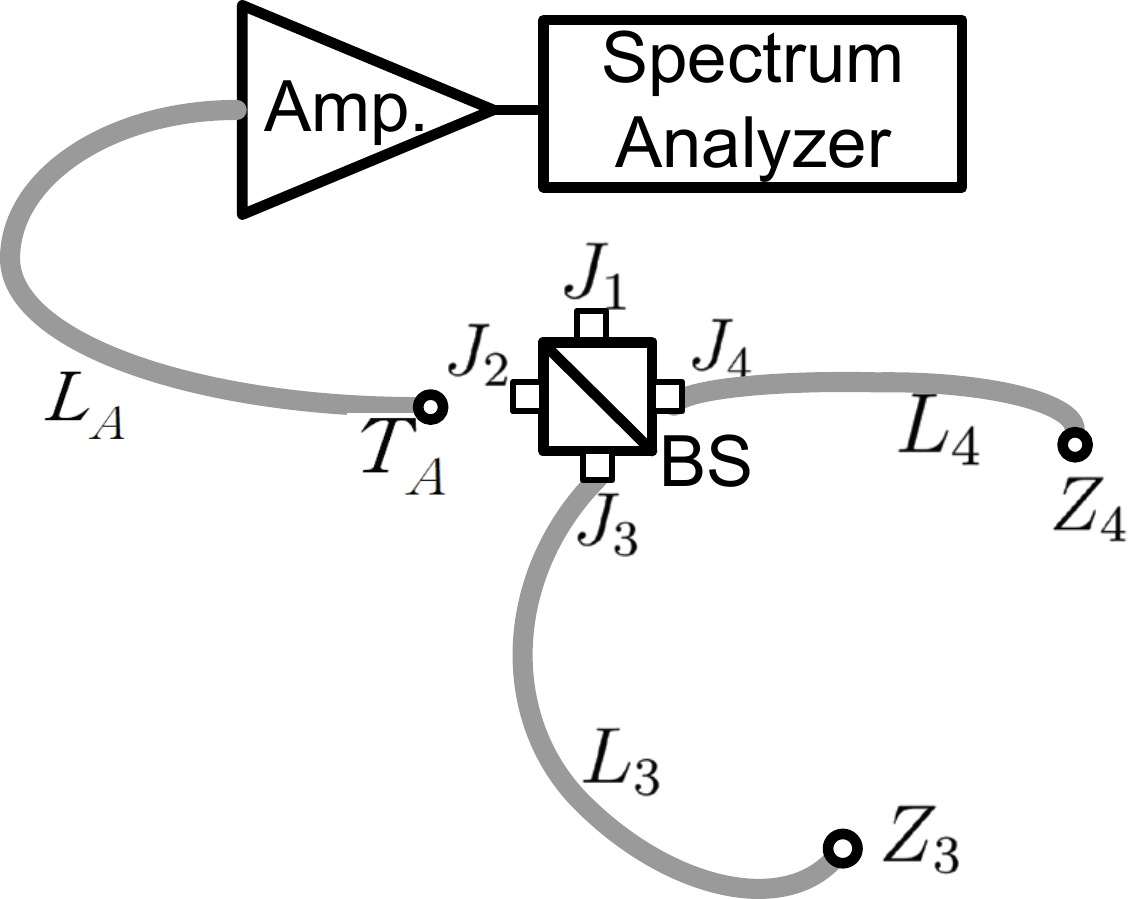}
\caption{Schematic of the noise measurement setup for a BNC cable. Noise amplified by a pre-amplifier is measured using a spectrum analyzer. The pre-amp is connected to a BNC cable of length $L_A$  with a terminal impedance $T_A$. The sum and difference of the electromagnetic waves entering through terminals $J_3$ and $J_4$ of the beam splitter are transmitted to terminals $J_1$ and $J_2$, respectively.
  $J_3$ and $J_4$ are connected to BNC cables with terminal impedances $Z_3$ and $Z_4$, and lengths $L_3$ and $L_4$, respectively. } \label{FigSetup}
\end{figure}
\begin{eqnarray}
v_{1m} = v_0   e^{-i  k L_A}   \Gamma_{l}  e^{-i  k  L_A}, \label{voltage1m}
\end{eqnarray}

where $k$ is the wave number of the electromagnetic wave. The term $\Gamma_{l}  $   represents the reflection coefficient at the load ($\frac{Z_l - Z_0} {Z_l + Z_0}$).
Upon reaching the input, this wave encounters the impedance $Z_b$, leading to a further reflection. The voltage of this reflected wave ($v_{1p}$) is:

\begin{eqnarray}
v_{1p} = v_0   e^{-i  k  L_A} \Gamma_{l}   e^{-i  k  L_A} \Gamma_{b} . \label{voltage1m}
\end{eqnarray}
, where  The term $\Gamma_b$  is the reflection coefficient at the source end ($ \frac{ Z_b - Z_0}{ Z_b + Z_0}$).

At this stage, the total electromagnetic wave amplitude at the device input is the superposition of the initial wave and the first two reflected waves, 
$v_0 + v_{1m} + v_{1p} $. 

 However, $v_{1p}$  subsequently propagates along the line, undergoes reflection at $Z_l$, returns to the input, and is partially reflected again by $Z_b$. This process of repeated propagation and reflection continues indefinitely.  The resulting superposition of all these waves at the device input leads to the total voltage $v_s$ becomes

\begin{eqnarray}
v_{s} &=& v_0 + v_{1m} + v_{1p} +v_{2m} + v_{2p}+ ... \nonumber \\
         &=& v_0 \frac{e^{2 i k L_A } + \Gamma_l }{  e^{2 i k L_A} - \Gamma_l \Gamma_b } . \label{voltagevsall}
\end{eqnarray}

In the specific case where the device impedance matches the characteristic impedance of the transmission line $(Z_b = Z_0)$, the reflection coefficient at the source end becomes zero. This eliminates all subsequent reflections from the source end, significantly simplifying the expression for $v_s$.  

The magnitude squared of this simplified voltage, representing the  power delivered to the device, is then:
\begin{eqnarray}
v^2 _{s} =  v_b ^2  \frac{Z_l ^2 \cos^2 ( k L_A ) + Z_0 ^2 \sin^2 (k L_A) }{ (Z_0 + Z_l )^2}
\label{voltageSQ}
\end{eqnarray}

The above expression can be recast in terms of the microwave frequency ($f$). The wavenumber ($k$) is related to the frequency by 
$k =  \frac{2\pi   f n }{c} $, where $c$  is the speed of light in vacuum  and $n$ is  the refractive index of the transmission line material:
\begin{eqnarray}
v^2 _{s}(f)  =  v_b ^2  \frac{Z_l ^2 \cos^2 ( 2 \pi f n  L_A /c ) + Z_0 ^2 \sin^2 (2 \pi f n  L_A /c ) }{ (Z_0 + Z_l )^2}
\label{voltageSQf}
\end{eqnarray}

Prior to incorporating the beamsplitter (BS) (see Figure \ref{FigSetup}), we conducted experiments varying both the length and termination impedance of the BNC cable connected to the pre-amplifier.  Measurements were performed using an E4443A spectrum analyzer and a SR445  pre-amplifier at room temperature. The measured electromagnetic signal originated from thermal noise generated by the 50-ohm input impedance of the pre-amplifier.  Cable lengths of 4 m and 8 m were tested, with the cable terminated in either a short circuit (0 $\Omega$) or an open circuit (infinite impedance).

\begin{figure}[h!]
    \centering
 \includegraphics[width=0.8\textwidth]{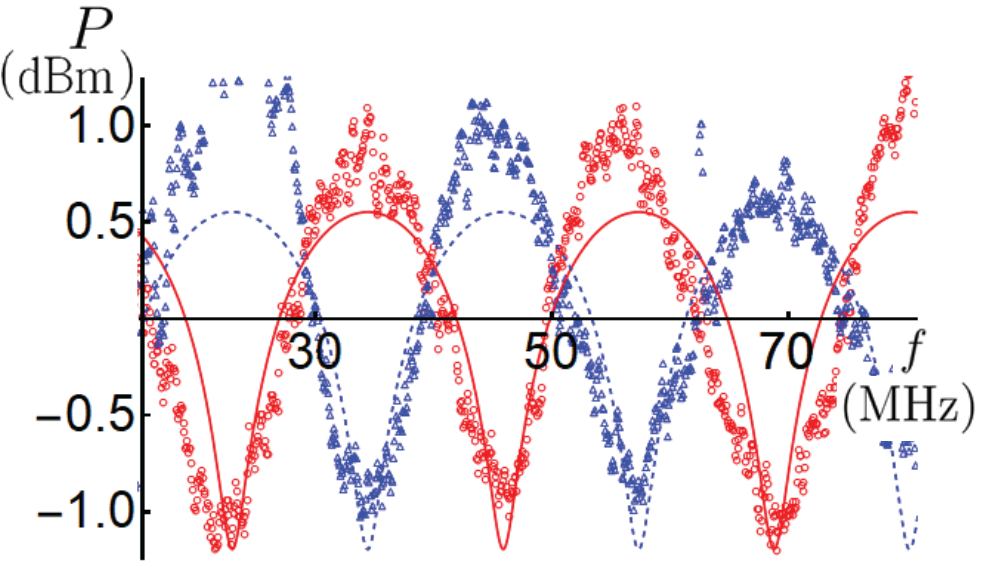}
\caption{Noise spectra measured using a spectrum analyzer, with the input BNC cable of the pre-amplifier set to a length of 4 m. Red circles and blue triangles indicate measurements taken with the cable end impedance set to 0 $\Omega$ and infinity, respectively. Values are normalized to the case where the cable end impedance is 50 $\Omega$. Theoretical predictions (solid red and dashed blue lines) were calculated using a BNC cable with a length of 4.08 m and a refractive index of 1.60.}
 \label{FigFour1}
\end{figure}

Figure \ref{FigFour1} presents the measured noise power as a function of frequency for a 4-m BNC cable, comparing the cases of short-circuit and open-circuit termination against a 50 $\Omega$ matched termination.  Red circles represent the short-circuit termination (0 $\Omega$), while blue triangles represent the open-circuit termination (infinite impedance). The solid and dashed curves represent fits to the data, incorporating the pre-amplifier gain and system errors.  The fitting function used was $ f = a + log(v_s^2 + sn) $ , where $a$, $sn$, and the cable length were fitting parameters.  Optimization yielded $a = -1.754$, $ sn = 1.91$, a cable length of 4.08 m, and a refractive index $n = 1.60$. As evident in Figure \ref{FigFour1}, the measured noise power exceeds the fitted values in regions of high noise. Conversely, in low-noise regions, the measured data aligns well with the theoretical predictions.

\begin{figure}[h!]
    \centering
 \includegraphics[width=0.8\textwidth]{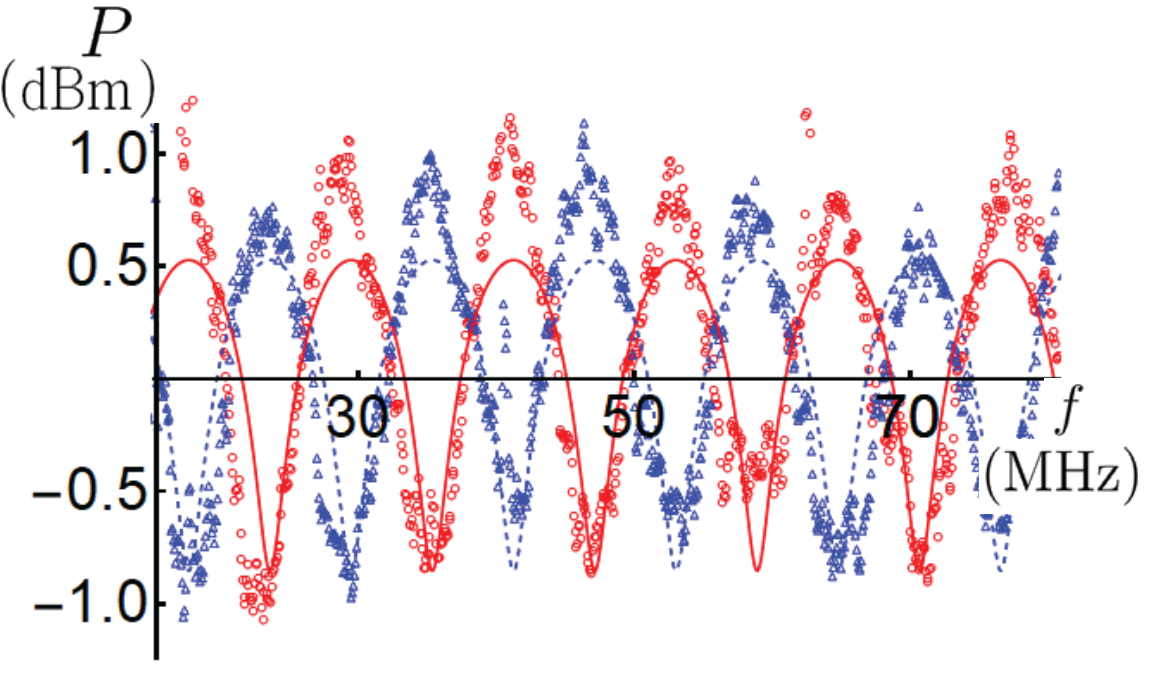}
\caption{Noise spectra measured using a spectrum analyzer, with the input BNC cable of the pre-amplifier set to a length of 8 m. Red circles and blue triangles indicate measurements taken with the cable end impedance set to 0 $\Omega$ and infinity, respectively. Values are normalized to the case where the cable end impedance is 50 $\Omega$. Theoretical predictions (solid red and dashed blue lines) were calculated using a BNC cable with a length of 7.93 m and a refractive index of 1.60..}
 \label{FigEight1}
\end{figure}

Figure \ref{FigEight1} displays analogous results for an 8-m BNC cable, again comparing short-circuit (red circles) and open-circuit (blue triangles) terminations to a 50-ohm matched load. The solid and dashed curves represent fits using the same values for $a$, $ sn$, and $n$ as in Figure  \ref{FigFour1}, with only the cable length adjusted to 7.93 m for optimal fitting.

In this section, we experimentally investigate the spatial distribution of electromagnetic waves associated with thermal noise by directly connecting a BNC cable to the preamplifier and adjusting the impedance at the cable end to induce reflections of the propagating waves. The measured spatial profiles of the thermal noise show good agreement with theoretical predictions for cable lengths of 4 m and 8 m. Considering that thermal noise does not generate new spatial modes but rather excites existing electromagnetic modes, the spatial structure of thermal noise reflects the spatial characteristics of these modes, which, in turn, correspond to the spatial distribution of vacuum fluctuations. Therefore, our findings provide experimental insight into the spatial profile of vacuum fluctuations.


\section{Thermal Noise Distribution in Beam-Splitter Configurations: Inferring Vacuum Fluctuation Suppression} \label{Results}

We analyze the propagation of electromagnetic waves and the contribution of thermal noise within a system incorporating a H2979 beam splitter (BS).  
The BS (Beam Splitter) in Figure \ref{FigSetup} combines electromagnetic waves input at ports $J_3$ and $J_4$ to produce a sum signal output at port $J_1$ and a difference signal output at port $J_2$. A BNC cable of length $L_A$ is connected directly to port $J_2$ for connection to the pre-amplifier. Port $J_1$ is terminated with a 50 $\Omega$ impedance to prevent reflections. BNC cables of lengths $L_3$ and $L_4$ are connected to ports $J_3$ and $J_4$, respectively, with characteristic impedances $Z_3$ and $Z_4$.

%

An initial voltage, $v_{20}$, applied to the input of a pre-amplifier, propagates along a cable of length $L_A$ to port $J_2$ of the BS.  The wave is then split and directed towards ports $J_3$ and $J_4$. The portion of the wave exiting $J_4$ travels along a cable, is reflected by an impedance $Z_4$, and subsequently splits again, propagating towards $J_1$ and $J_2$. A similar process occurs for the wave exiting $J_3$, reflecting from impedance $Z_3$ and splitting towards $J_1$ and $J_2$. Port $J_1$ is terminated with a 50 $\Omega$ impedance, matching that of the cable, preventing further reflections at this point.  Therefore, the voltage at the pre-amplifier input resulting from the reflected waves via the BS can be expressed as:

\begin{align}
v_{2a} = v_{20} \left[ 1 +  e^{-i k L_A + i \omega \tau_{32}} s_{32} \Gamma_3 e^{-2 i k L_3} e^{+ i \omega \tau_{23}} s_{23}  e^{-i k L_A} \right. \nonumber \\
\left. + e^{-i k L_A +i  \omega \tau_{42}} s_{42} \Gamma_4 e^{-2 i k L_4} e^{ i \omega \tau_{24}} s_{24}  e^{-i k L_A} \right],
\label{voltageV2a}
\end{align}

where $\Gamma_i = (Z_i - Z_0)/(Z_i + Z_0)$ represents the reflection coefficient at impedance $Z_i$, and $Z_0$ is the characteristic impedance of the cable, $\omega$ is the angular frequency,  $s_{ij} $ is the S-matrix of the BS and $\tau_{ij} $ represents the propagation delay between ports $i$ and $j$ of the BS.

If the thermal noise generated by the resistor connected to the pre-amplifier is reflected back to the pre-amplifier through the region coupled via the beam
 splitter and the associated cables, as described in Eq.\ref{voltageV2a}, the thermal noise originating from the resistors located 
 at the termination of the cables connected to the beam splitter is transmitted to the pre-amplifier as given by Eq. \ref{voltageOut},
  which is derived in detail in the Supplementary Materials. 
  This expression can be simplified by considering the limiting cases where $Z_3$ and $Z_4$ approach zero or infinity, yielding the following result.

\begin{eqnarray}
v ^2 (m_3 ,m_4 ) &=& \frac{4 k_b T Z_0}{8} \{4-2 (-1)^{m_3} \cos(2 (kL_A + kL_3 -\omega \tau_{23} ) ) \nonumber \\ 
&-& 2 (-1)^{m_4} \cos(2 (kL_A+kL_4-\omega \tau_{24})) 
\nonumber \\
&-& (-1)^{m_3+m_4} \cos (2 kL_3 -2 kL_4-\omega \tau_{13}+\omega \tau_{14}-\omega \tau_{23}+\omega \tau_{24})\nonumber  \\ 
&+&(-1)^{m_3 +m_4 } \cos(2 (kL_3-kL_4-\omega \tau_{23}+\omega \tau_{24}))\}
\label{voltageSim}
\end{eqnarray}

Here, $m_i$ is an index representing the state of impedance $Z_i$, taking the value 0 when $Z_i = 0$ and 1 when $Z_i \rightarrow \infty$.

\begin{figure}[h!]
    \centering
 \includegraphics[width=0.8\textwidth]{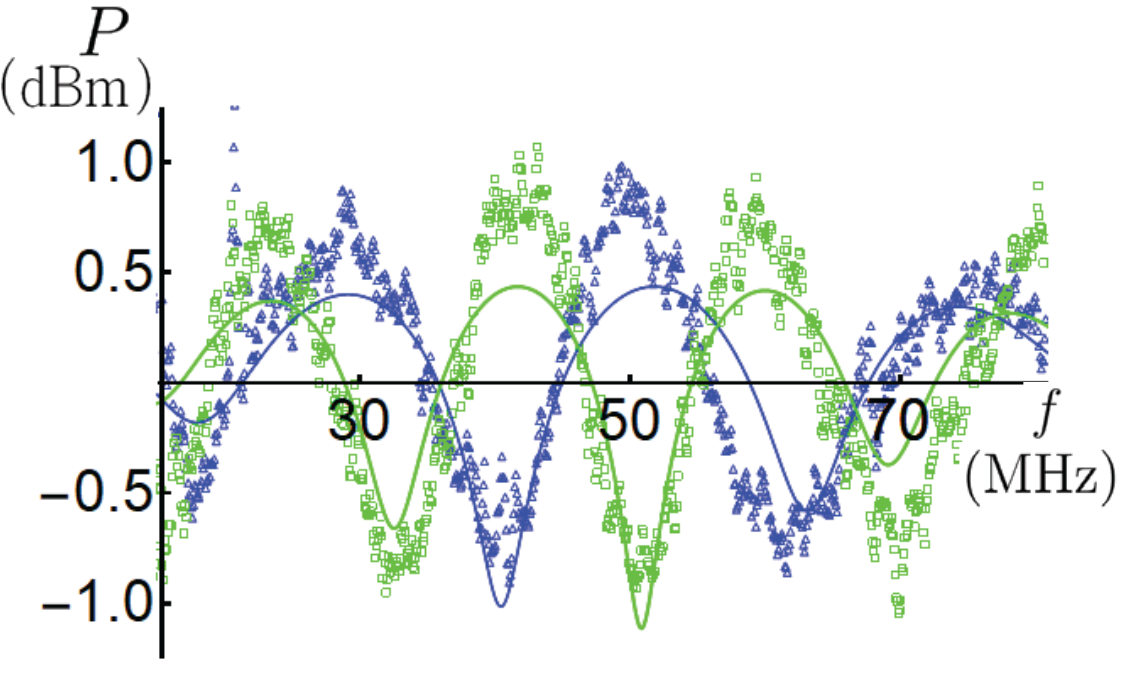}
\caption{Noise characteristics measured after the electromagnetic waves combined at the BS (Beam Splitter) were amplified by the pre-amplifier. A BNC cable with a length of 2 m and short-circuited impedance (0)  was connected to terminal $J_3$  of the BS, while a BNC cable with a length of 1 m and an open-ended impedance (infinity) was connected to terminal $J_4$. The noise characteristics were measured for cases where the BNC cable length between the pre-amplifier and the BS was 2 m and 1 m, represented by green squares and blue triangles, respectively. Theoretical values calculated for cable lengths of 1.98 m and 0.98 m are shown as green and blue solid lines, respectively.  } \label{FigbsTwo1}
\end{figure}

Figure \ref{FigbsTwo1} presents noise spectra measured at a spectrum analyzer connected to output port $J_2$ of a beam splitter (BS) via a pre-amplifier, with specific cabling and termination conditions at the BS input and output ports. Input port $J_3$ is connected to a 2-meter BNC cable terminated with a 0
 $\Omega$ impedance, while input port $J_4$ is connected to a 1-meter BNC cable terminated with an open circuit (infinite impedance). Output port $J_1$ is terminated with a 50  $\Omega$ impedance, matching the cable impedance. The green squares represent data acquired with a 2-meter cable between $J_1$ and the pre-amplifier, and the blue triangles represent data with a 1-meter cable in the same location. The data are presented as the noise power relative to that measured when the pre-amplifier input is terminated with a 50  $\Omega$ load. The solid green line is a fit to the data using Equation \ref{voltageSim} with the following parameters: $\tau_{41} = \tau_{31} = 5.31$ ns, $\tau_{42} = \tau_{32} = 8.46$ ns, $a = -1.27$, $s_n = 1.10$, and a distance of 0.98 m between the pre-amplifier and $J_1$. The solid blue line represents a fit with identical parameters, except for an increased pre-amplifier to $J_1$ distance of 1.98 m. The theoretical model shows good agreement in the low-noise regions, while discrepancies in high-noise regions are likely attributable to the intrusion of external noise sources.

\begin{figure}[h!]
    \centering
 \includegraphics[width=0.8\textwidth]{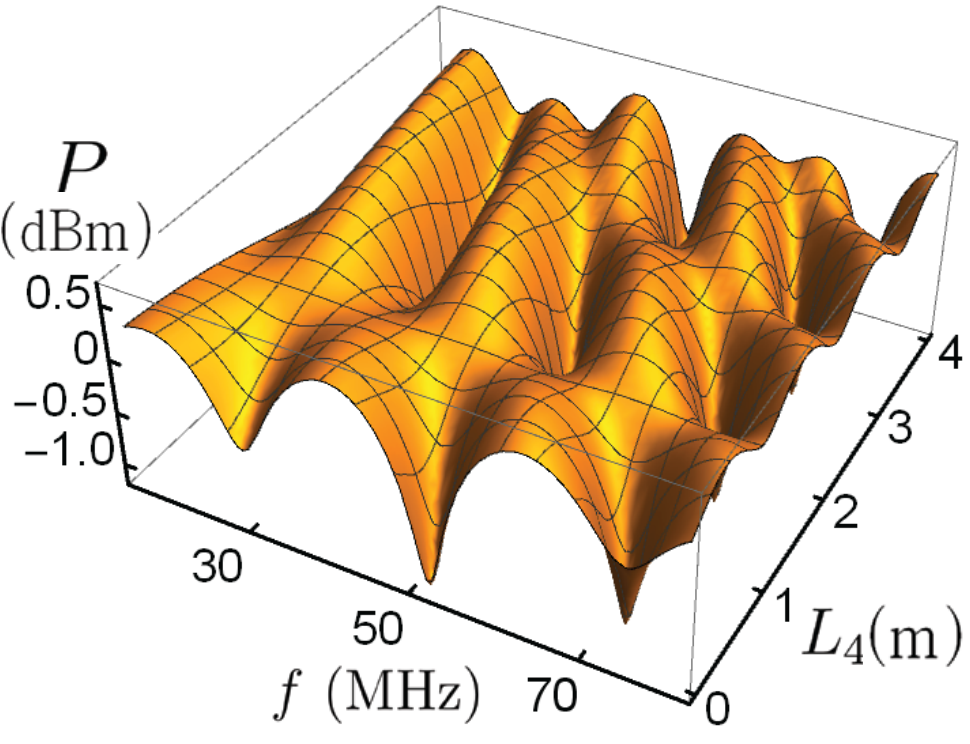}
\caption{Theoretical noise power as a function of frequency and BNC cable length at one input of a beam splitter (BS). The calculation assumes a 0 $\Omega$ impedance termination at input port $J_4$ (variable length cable). Input port $J_3$ has a fixed 1 m BNC cable terminated with infinite impedance (open circuit). The pre-amplifier is connected to the BS via a 2 m BNC cable.} \label{Figbs3D1}
\end{figure}

To predict the characteristics of the thermal noise measured at the spectrum analyzer, as a function of BNC cable length at the BS input, theoretical calculations are presented in Figure \ref{Figbs3D1}. The length of cable between $J_2$ and pre-amplifier is 2m. The $J_1$ is terminated with 50 $\Omega$. A 1-meter BNC cable, terminated with an open circuit (infinite impedance), was connected to input port $J_3$. Another cable terminated to 0  $\Omega$ was connected to the other input port $J_4$. Calculations were performed for the noise power spectral density as a function of frequency, varying the length ($L_4$) of a BNC cable terminated in a short circuit (0  $\Omega$) connected to input port $J_4$ from 0 to 4 meters. The calculations reveal complex interference patterns arising from reflections of electromagnetic waves within the cables connected to the BS input ports.

\begin{figure}[h!]
    \centering
 \includegraphics[width=0.8\textwidth]{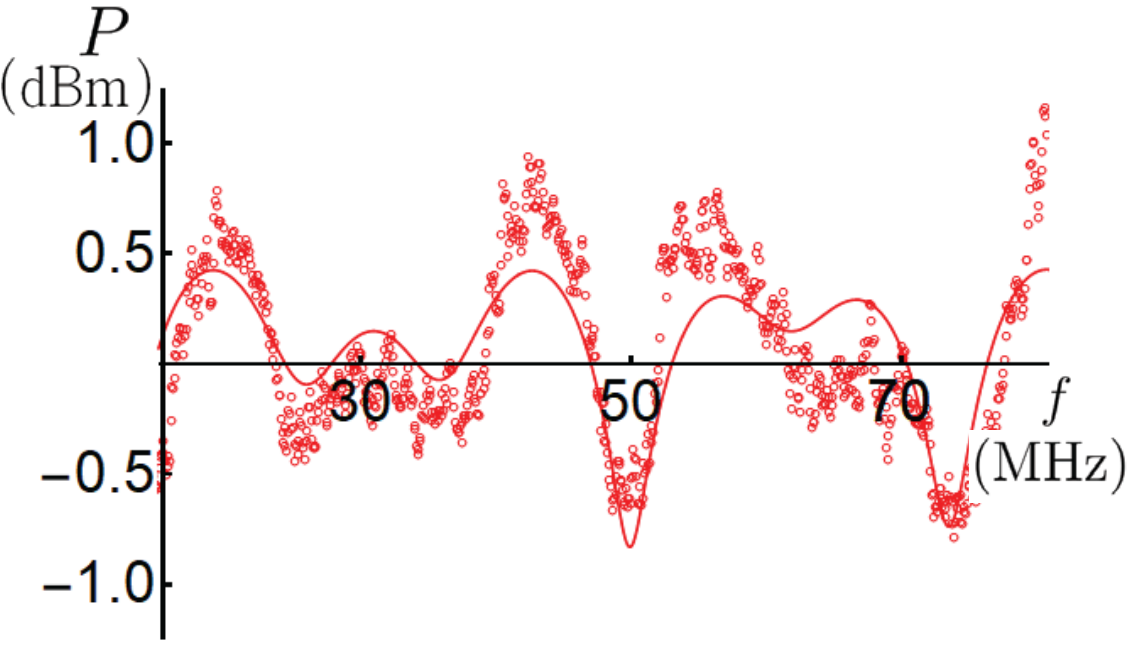}
\caption{Noise characteristics measured after the electromagnetic waves combined at the BS (Beam Splitter) were amplified by the pre-amplifier. A BNC cable with a length of 1 m and an open-ended impedance (infinity) was connected to terminal $J_3$ of the BS, while a BNC cable with a length of 4 m and a short-circuited impedance (0) was connected to terminal $J_4$. The BNC cable length between the pre-amplifier and the BS was 2 m. Theoretical values calculated for cable lengths of 1.98 m is shown as red solid line.} \label{FigbsFour1}
\end{figure}

Figure \ref{FigbsFour1} shows the calculated thermal noise power spectral density (red circles) for the case of $L_4$ = 4 m in Figure \ref{Figbs3D1}. The red solid line is a fit to this calculated data, using the same parameters as in Figure \ref{FigbsTwo1}, but with a pre-amplifier to $J_1$ distance of 1.98 m and a cable length of 3.98 m connected to $J_4$. The data in the lower noise regions are in strong agreement with the theoretical fit.

In this section, we experimentally investigate the spatial distribution of electromagnetic waves associated with thermal noise in several modes combined by the BS. The measured spatial profiles of the thermal noise show good agreement with theoretical predictions.  These results prove the spatial dependence of the vacuum fluctuation in the BS system with a mirror.

\section{Discussion} \label{discussion}

 In our experiment, we present data that reveal the spatial distribution of thermal noise propagating through BNC cables, particularly those connected to a beam splitter. While such results could, in principle, be interpreted by calculating the impedance at the preamplifier (pre-amp) input based on microwave theory, our approach focuses instead on how thermal noise generated at the 50-ohm resistor at the pre-amp input couples into the electromagnetic modes supported by the external environment, including the BNC cables and the beam splitter.

 In essence, our measurements probe the electromagnetic modes that can be formed within the spatial region defined by the BNC cables. The observed mode structure inherently reflects the behavior of the electromagnetic modes that would exist in the absence of any thermal noise. In this sense, what we are measuring is the spatial distribution of vacuum fluctuations — the quantum noise that remains even when thermal noise is absent — as shaped by the boundary conditions imposed by the experimental setup.

 Therefore, by measuring the spatial distribution of thermal noise originating from the 50-ohm resistor at the pre-amp input, we indirectly characterize the electromagnetic mode structure within the BNC cables. This, in turn, provides a direct window into the spatial distribution of vacuum fluctuations within those modes. For example, when the far end of the BNC cable is terminated with a short (0 ohm), the electromagnetic waves are reflected in a manner analogous to reflection from a mirror. This allows us to infer how vacuum fluctuations are modified in the vicinity of a reflective boundary.

 Does this merely represent a reinterpretation of a known phenomenon? We argue that the results provide a more precise understanding of vacuum fluctuations, with potential implications for quantum-limited precision measurements such as gravitational wave detection, where squeezed states are employed to suppress noise below the vacuum level. In such systems, being able to shape the spatial properties of vacuum fluctuations via mirrors and beam splitters could offer a new degree of control.

 While immediate application in the optical regime may be limited by the practical difficulty of achieving sub-wavelength spatial resolution in detectors, our findings suggest alternative applications. For instance, in intensity-stabilized light sources, a fraction of the output is usually measured and fed back to maintain constant power. In this process, vacuum noise entering through the unused port of a beam splitter limits the ultimate stability of the source \cite{Youn1994}. By inserting a mirror at this unused port, it becomes possible to suppress vacuum noise of specific phases, thus enabling feedback stabilization below the standard quantum limit. This approach is particularly promising in the radio-frequency domain, where current technology already allows implementation.

 In conclusion, our experiment demonstrates the spatial distribution of thermal noise in the radio-frequency domain in the presence of reflective boundaries such as mirrors, allowing us to infer the corresponding vacuum mode structure. These findings are broadly applicable across the electromagnetic spectrum and suggest that by incorporating mirrors in beam splitter configurations, one can stabilize the intensity of electromagnetic sources by suppressing specific components of quantum noise below the standard quantum limit.
 

\clearpage
\section*{Supplementary Materials}

To evaluate the propagation of thermal noise generated at the termination resistances of the cables connected to ports $J_3$, $J_4$, and $J_1$ 
of the beam splitter to the input of the preamplifier, we begin by analyzing the contribution from the resistance $Z_3$ located at the end of the cable connected to port $J_3$. The resulting thermal noise contribution is given by:

\begin{align}
v_{3a} =  v_{30} e^{-i k L_3 + i \omega \tau_{23}} s_{23}  e^{-i k L_A},
\label{voltageV3a}
\end{align}

where $v_{30} = \frac{Z_0}{Z_3 + Z_0}  \sqrt{4 k_B T Z_3}$, representing the voltage noise spectral density at the source, with $k_B$ being Boltzmann's constant and $T$ the temperature.

Similarly, the thermal noise contribution from impedance $Z_4$ is:

\begin{align}
v_{4a} =  v_{40} e^{-i k L_4 + i \omega \tau_{24}} s_{24}  e^{-i k L_A},
\label{voltageV4a}
\end{align}

where $v_{40} = \frac{Z_0}{Z_4 + Z_0}  \sqrt{4 k_B T Z_4}$.

Finally, thermal noise originating from the termination impedance at $J_1$ propagates through the BS, with portions reaching the pre-amplifier input via $J_2$.  This contribution is described by:
\begin{align}
v_{1a} = v_{10}  e^{-i k L_1} \left[ e^{+ i \omega \tau_{31}} s_{31} \Gamma_3 s_{23}  e^{- i k L_A + i \omega \tau_{23}} \right. \nonumber \\
\left. + e^{ i \omega \tau_{41}} s_{41} \Gamma_4 s_{24}  e^{- i k L_A + i \omega \tau_{24} } \right],
\label{voltageV1a}
\end{align}

where $v_{10} = \frac{Z_0}{Z_1 + Z_0}  \sqrt{4 k_B T Z_1}$.

The S-matrix parameters characterizing the BS are: $s_{32}=s_{23} = -1/\sqrt{2}$ and $s_{24}=s_{42}=s_{31}=s_{13}=s_{14}=s_{41}= 1/\sqrt{2}$. The phase shift $w_{ij}$ represents the propagation delay between ports $i$ and $j$ of the BS.

Considering the statistically independent nature of thermal noise sources, the total mean-squared voltage at the pre-amplifier input is:

\begin{eqnarray}
v_{sum}^2  &=& \frac{4 k_b T Z_0}{8 (Z_0+z_2)^2 (Z_0+Z_3)^2 (Z_0+Z_4)^2} \nonumber \\
&\times&  
(Z_0+z_2)^2 \{ Z_0^4+Z_3 ^2 Z_4 ^2 + 4 Z_0 ^3 (Z_3 + Z_4 )+ 4 Z_0 Z_3 Z_4 (Z_3 +Z_4 ) +
 Z_0 ^2 (Z_3 ^2 +12  Z_3 Z_4 + Z_4 ^2)    \nonumber \\ 
 &  -&  (Z_0^2-Z_3^2)
  (Z_0^2-Z_4^2) \cos (2 kL_3 -2
   kL_4-\omega \tau_{13}+\omega \tau_{14}-\omega \tau_{23}+\omega \tau_{24}) \} \nonumber \\
&+&
   4 Z_0 z_2 \{  3 Z_0 ^4 + 4 Z_0 ^3 Z_3 + 3 Z_0 ^2 Z_3 ^2 + 4 Z_0 ^3 Z_4 + 4 Z_0 ^2 Z_3 Z_4 + 4 Z_0 Z_3 ^2 Z_4  \nonumber \\ &+&
   3 Z_0 ^2 Z_4 ^2 + 4 Z_0 Z_3 Z_4 ^2 + 3 Z_3 ^2 Z_4 ^2  \nonumber \\
      & -& 2 (Z_0^2-Z_3^2) (Z_0+Z_4)^2 \cos (2
   (kL_A+kL_3-\omega \tau_{23}))\nonumber \\  &-& 2 (Z_0+Z_3)^2 (Z_0^2-Z_4^2)
   \cos (2 (kL_A+kL_4-\omega \tau_{24})) \nonumber \\ 
   &+& Z_0^4 \cos (2
   (-kL_3+kL_4+\omega \tau_{23}-\omega \tau_{24}))-Z_0^2 Z_3^2 \cos (2
   (-kL_3+kL_4+\omega \tau_{23}-\omega \tau_{24})) \nonumber \\ &-& Z_0^2 Z_4^2 \cos (2
   (-kL_3+kL_4+\omega \tau_{23}-\omega \tau_{24})) \nonumber \\ &+& Z_3^2 Z_4^2 \cos (2
   (-kL_3+kL_4+\omega \tau_{23}-\omega \tau_{24}))\}
\label{voltageOut}
\end{eqnarray}


\begin{thebibliography}{7}

\bibitem{casimir1948}
H. B. G. Casimir, D. Polder, The influence of retardation on the London-van der Waals forces. \textit{Phys. Rev.} \textbf{73}, 360 (1948).

\bibitem{jae1980}
W. Jhe, A. Anderson, E. A. Hinds, D. Meschede, L. Moi, S. Haroche, Suppression of spontaneous decay at optical frequencies: Test of vacuum-field anisotropy in confined space. \textit{Phys. Rev. Lett.} \textbf{58}, 666--669 (1987).

\bibitem{youn1995}
S.-H. Youn, J.-H. Lee, J.-S. Chang, Modulation of the vacuum field in the beam splitter and mirror system. \textit{Opt. Quant. Electron.} \textbf{27}, 355 (1995).

\bibitem{wadood2019}
S. A. Wadood, J. T. Schultz, A. N. Vamivakas, C. R. Stroud Jr., Do remote boundary conditions affect photodetection? \textit{J. Mod. Opt.} \textbf{66}, 1116--1123 (2019).

\bibitem{pozar2011}
D. M. Pozar, \textit{Microwave Engineering}, 4th edn (Wiley, 2011).

\bibitem{shahmoon2017}
E. Shahmoon, Casimir forces in transmission-line circuits: QED and fluctuation-dissipation formalisms. \textit{Phys. Rev. A} \textbf{95}, 062504 (2017).

\bibitem{Youn1994}
S.-H. Youn, W. Jhe, J.-H. Lee, J.-S. Chang, Effect of vacuum fluctuations on the quantum-mechanical amplitude noise of a laser diode in feedback systems. \textit{J. Opt. Soc. Am. B} \textbf{11}, 20--26 (1994).

\end{thebibliography}
\end{document}